\documentclass[onecolumn,showpacs,aps,prd]{revtex4}
\def\OL{\overline}

\usepackage{graphicx}
\usepackage{dcolumn}
\usepackage{amsmath}
\usepackage{epsfig}
\RequirePackage{xspace}

\begin{document}

\def\op{{\cal O}}
\def\lsim{\mathrel{\lower4pt\hbox{$\sim$}}\hskip-12pt\raise1.6pt\hbox{$<$}\;}
\def\gsim{\mathrel{\lower4pt\hbox{$\sim$}}\hskip-10pt\raise1.6pt\hbox{$>$}\;}
\def\Dd{\psi}
\def\pp{\lambda}
\def\ket{\rangle}
\def\bra{\langle}
\def\BAR{\bar}
\def\xba{\bar}
\def\fa{{\cal A}}
\def\fm{{\cal M}}
\def\fl{{\cal L}}
\def\ufs{\Upsilon(5S)}
\def\ufour{\Upsilon(4S)}
\def\xcp{X_{CP}}
\def\ynotcp{Y}
\vspace*{-.5in}
\def\bfb{{\bf B}}
\def\fd{r_D}
\def\fb{r_B}
\def\hatA{\hat A}
\def\hatfd{{\hat r}_D}
\def\D{{\bf D}}
\def\pcc{(+ charge conjugate)}

\def\KS{K_S}
\def\KL{K_L}
\def\dmd{\Delta m_d}
\def\dms{\Delta m_s}
\def\dgd{\Delta \Gamma_d}
\def\dgs{\Delta \Gamma_s}

\def\uglu {\hskip 0pt plus 1fil minus 1fil}
\def\uglux{\hskip 0pt plus .75fil minus .75fil}

\def\slashed#1{\setbox200=\hbox{$ #1 $}
    \hbox{\box200 \hskip -\wd200 \hbox to \wd200 {\uglu $/$ \uglux}}}

\def\slpar{\slashed\partial}
\def\sla{\slashed a}
\def\slb{\slashed b}
\def\slc{\slashed c}
\def\sld{\slashed d}
\def\sle{\slashed e}
\def\slf{\slashed f}
\def\slg{\slashed g}
\def\slh{\slashed h}
\def\sli{\slashed i}
\def\slj{\slashed j}
\def\slk{\slashed k}
\def\sll{\slashed l}
\def\slm{\slashed m}
\def\sln{\slashed n}
\def\slo{\slashed o}
\def\slp{\slashed p}
\def\slq{\slashed q}
\def\slr{\slashed r}
\def\sls{\slashed s}
\def\slt{\slashed t}
\def\slu{\slashed u}
\def\slv{\slashed v}
\def\slw{\slashed w}
\def\slx{\slashed x}
\def\sly{\slashed y}
\def\slz{\slashed z}
\def\slE{\slashed E}

\renewcommand{\Re}{\ensuremath{{\rm Re}}}
\renewcommand{\Im}{\ensuremath{{\rm Im}}}

\def\cals{{\cal S}}

\title{

\begin{flushright}
{BNL-HET-06/15~~~}\\
\end{flushright}

\vskip 10mm
\large\bf
\boldmath
Clean Signals of CP-violating and CP-conserving New Physics in
$B \to P V \gamma$ Decays at $B$ Factories and Hadron Colliders.
\begin{center}
\end{center}
}

\author{David Atwood}
\affiliation{
  Department of Physics and Astronomy, Iowa State University, Ames, IA 50011
}
\author{Tim Gershon}
\affiliation{
  Department of Physics,
  University of Warwick,
  Coventry, CV4 7AL, UK
}
\author{Masashi Hazumi}
\affiliation{
  High Energy Accelerator Research Organization (KEK), Tsukuba, Ibaraki, Japan
}
\author{Amarjit Soni}
\affiliation{
  Theory Group, Brookhaven National Laboratory, Upton, NY 11973
}

\date{\today}

\begin{abstract}


In radiative $B$ decays to final states containing 
one spin zero meson, one spin one meson and a photon,
the photon polarization can be measured 
from the angular distribution of the final state. 
The P-odd, C-even components of this distribution lead to 
triple correlation asymmetries that are very sensitive to new physics 
as they are likely to receive only tiny
contributions from the Standard Model.
There are also (CP-conserving) forward-backward asymmetries 
to which there may be SM contributions of a few percent;
nevertheless there is a data driven means to overcome the SM ``pollution''. 
These latter observables can be used to search for a class of New Physics 
which does not necessarily possess a new CP-odd phase,
and wherein the current structure is different from 
the left-handed electroweak theory of the Standard Model. 
The mode $B^\pm \to K^\pm \phi \gamma$ is particularly useful due to the
relatively large branching ratio and its distinctive final state but
there are dozens of such final states in the decays of 
$B_u$, $B_d$ and $B_s$ mesons 
where the analysis is applicable. 
In many cases,
after the decay of the spin one resonance,
several of these channels lead to only charged mesons and a photon 
in the final state, so they may well be accessible in a hadronic environment 
in addition, of course, to $e^+ e^-$ (Super) $B$ Factories.
In passing we also briefly explain why the CP-conserving forward-backward 
asymmetries in $B \to P V \gamma$ decays is a more reliable signal
of new physics compared to the (CP-conserving) 
transverse polarization in  $B \to V_1 V_2$ modes even though, of course,
the latter may be somewhat more abundant.        
\end{abstract}


\maketitle

%
%
%
%

\section{Introduction}
%
In the past few years, the spectacular successes of the two
asymmetric $B$ factories, have helped us attain  an important
milestone in our understanding of CP violation~\cite{mh_ichep06}.
It has been established that the CKM-paradigm~\cite{ckm}
provides the dominant contribution
to the observed CP violation in the $B$ and $K$ decays. 
While essentially compelling theoretical arguments suggest that beyond
the SM source(s) of CP violation must exist, the $B$ factories results
strongly imply that the effects of these new CP-odd phase(s) in
$B$ physics can only be a subdominant perturbation. 
Future experimental efforts will therefore clearly require large samples of
clean $B$ mesons, that is a Super $B$ Factory~\cite{super-b}.
Hadron colliders of course produce impressive numbers of $B$ mesons and baryons. 
Though harnessing these is a non-trivial challenge, 
the CDF and D0 experiments at the Tevatron 
have already made significant contributions to $B$ physics
through their successful determinations of 
the $B_s$ mass-difference~\cite{cdf_d0}.
With the arrival of LHC and LHCb, in the next year or two, 
we can anticipate many more important inputs on $B$ physics. 
To facilitate searches of the expected small effects 
from beyond the Standard Model (BSM) physics, 
theory can help greatly by developing a class of 
clean null tests of the SM~\cite{gs06}.
These pertain to observables for which the SM
predictions are vanishingly small. Motivated by these considerations
we continue to explore the use of photon polarizations in exclusive
radiative $B$ decays to search for New Physics (NP).

In this paper we focus on final states in $B$ decays consisting
of a spinless (say pseudoscalar $P$) meson in addition to 
a spin one (say vector $V$) meson and the photon 
({\it i.e.} $B \to PV\gamma$). 
We will show that 
therein the SM makes an unambiguous prediction that a class of  
{\it direct} CP violation asymmetries must be vanishingly small, 
{\it i.e.} $\lsim 1\%$, 
yielding a type of ``golden'' observable that can be used for 
searching for BSM sources of CP violation. 
In addition, $PV\gamma$ final states are also very
powerful in that we are able to construct (CP-conserving)
forward-backward asymmetries involving photon polarization for which
the SM pollution is predicted to be less than a few percent. 
Therefore, these latter observables can be very useful 
in searching for NP which is not necessarily endowed with a new CP-odd phase. 
Indeed, these methods based on using the photon polarization, 
represent an especially sensitive means to search for BSM sources of 
right-handed currents~\cite{ags97}.
Since $B_u$, $B_d$, $B_s$ can all be used, 
altogether there are dozens of decays modes 
where this formalism can be applied; see Table~\ref{many}. 
Furthermore, several of the final states appear accessible 
in a hadron collider experiment,
since 
after the decay of the vector resonance,
the final state consists of only charged mesons and an energetic photon.

It is useful to compare and contrast the use of $PV\gamma$ final states
with hadronic $B \to V_1 V_2$ decays as a tool for searching for new physics. 
In this regard, we believe even though $PV\gamma$ decays are somewhat rarer,
the presence of
the photon introduces considerable simplification and allows us to make 
more reliable predictions, based on the SM, 
especially with regard to CP-conserving forward-backward asymmetries, 
in contrast to the case of the expected size of
(CP-conserving) transverse polarization in $V_1 V_2$ final states. 
The reason  for this simplification can be traced to the fact
that for radiative decays the relevant effective Hamiltonian 
consists of a collection of 2-quark operators 
whereas for the purely hadronic final states ($V_1 V_2$) 
it is built on 4-quark operators. 
While the hadronic matrix elements for the 2-quark
operators are quite cumbersome, 
the case of the 4-quark operators is significantly more complicated. 
Therein issues of enhanced chiral corrections, 
weak annihilation~\cite{pqcd, bn_s4}, 
charming penguins~\cite{romans}, final state rescattering effects~\cite{ccs},
penguin induced annihilation~\cite{ak,da_prl}, {\it etc.} 
render reliable predictions extremely difficult and quite out of reach.   


The paper is organized as follows. 
Section II provides a more detailed introduction,
highlighting the motivation to study the modes under discussion.
Section III discusses
how the photon polarization parameters can be determined
through the observables accessible to $PV\gamma$ modes.
In section IV we use an illustrative toy model
which has a spin one resonance saturating the $PV$ hadrons to discuss
the various observables and asymmetries. 
Section V describes how to test whether the number of partial waves used
gives an adequate account of the data. 
Section VI and VII deal briefly with $B_d$ and $B_s$ respectively. 
Section VIII briefly discusses SM backgrounds 
and in section IX we give a brief summary.

\section{Motivation}

As is well known inclusive radiative decays ($B \to X_s \gamma$)
have for a long time provided strong constraints on many models of NP. 
The theoretical calculation for this is by now highly matured;
in fact just recently it has been advanced~\cite{mm06}
to NNLO with the SM prediction, 
${\cal B}(B \to X_s \gamma$, $E_\gamma \gsim 1.6 \ {\rm GeV}) =
(3.15 \pm 0.23) \times 10^{-4}$,
which is in good agreement with the latest 
experimental determination,
$(3.55 \pm 0.24 \pm 0.10 \pm 0.03) \times 10^{-4}$~\cite{btosgamma,hfag}.
While this is a very valuable test of the SM, 
since the inclusive branching ratio represents an incoherent sum of the
two polarization states of the photon, 
it is clearly not a sensitive test of one very important prediction of the SM: 
the photon in $b$ quark decays has to be predominantly left-handed (LH)
and those $\bar b$ quark  decays has to be predominantly right-handed (RH). 
This observation in 1997~\cite{ags97} led to, for the first time, a
proposal to use exclusive radiative decays to subject the SM to this
class of tests. This is particularly noteworthy since exclusive
final states are appreciably easier than the inclusive ones from the
experimental perspective, whereas ironically theoretical predictions
for exclusive modes are usually extremely challenging. It is,  
therefore, especially gratifying that exclusive radiative decays can
be used for precise tests of the SM following the ideas of~\cite{ags97}.

The key point of~\cite{ags97} is that the LH nature of the weak
interaction in the SM characteristically endows the outgoing photon
in $b$ quark decays to be essentially LH. 
Indeed, in the SM, the lowest order (LO) effective Hamiltonian $H_{\rm eff}$ 
implies that in $b$ quark decays the amplitude for emission of RH photons 
vanishes in the limit that the mass of the outgoing quark vanishes; 
{\it i.e.} for $b \to s\gamma$ the relevant quark mass 
is that of the strange quark ($m_s$) and for $b \to d\gamma$ it is $m_d$. 
This has the important consequence that in the SM,
mixing-induced (time-dependent) CP asymmetries, in channels such as
$B \to K^{*0} \gamma$, 
with $K^{*0}$ to a self-conjugate final state ($\KS \pi^0$), 
is severely suppressed.

Since QCD preserves chirality, emission and reabsorption of gluons
does not change the above predictions of the SM. However, a class of
higher order QCD corrections involving the emission of real gluons can
provide SM ``pollution'' as first emphasised in~\cite{grinstein1}. 
Of course this extra gluon must have extra suppression, 
in addition to being higher order in $\alpha_s$, 
to the extent that it has to find itself in the wave-function 
of the exclusive hadron (such as $K^{*0}$ in the above example). 
The relevant exclusive hadronic matrix elements are 
extremely difficult to estimate reliably. 
Using essentially a dimensional estimate,
Ref.~\cite{grinstein2} suggested that these higher order
contributions could be quite large $\approx {\cal O}(10\%)$, 
considerably bigger than what the LO $H_{\rm eff}$ gives. 
Recently two very useful explicit calculations 
of such higher order contributions have appeared. 
Ref.~\cite{matsumori} used pQCD and Ref.~\cite{bz} used QCD sum rules. 
In sharp contrast with~\cite{grinstein2}, 
both these works find the effect of the higher order corrections 
to be quite small and essentially confirm that the LO $H_{\rm eff}$ 
remains the dominant contribution.

Clearly it is quite difficult to reliably estimate, in a model
independent way, the effect of this SM ``pollution''. 
Fortunately, there are experimental avenues to bypass this theoretical
limitation~\cite{aghs}.  The experimental solutions require
higher luminosities but nonetheless, data driven solutions do exist.

For one thing, to the extent that the LO $H_{\rm eff}$ is relevant, 
the SM predicts that the time-dependent CP asymmetry
for related modes containing higher kaonic resonances, 
such as $B \to K^*(1270)\gamma$,  $B \to K^*(1410)\gamma$,
$B \to K_2^*(1430)\gamma$, {\it etc.}, 
is the same as that for $B \to K^*(892)\gamma$
(up to the charge conjugation eigenvalue of the final state in which
the resonance is reconstructed~\cite{aghs}). 
Indeed, since higher order gluonic contribution to the matrix elements for
the different kaonic resonances (with different masses or different
quantum numbers) are expected to be appreciably different, 
a dispersion in measured time-dependent CP asymmetries 
would indicate the presence of the higher order contributions. 
Conversely, to the extent that the asymmetries are the same, 
the results can be combined to enhance the precision.

Building on the work of~\cite{gh},
Ref.~\cite{aghs} made an important generalization to~\cite{ags97}, 
by emphasizing that the hadron accompanying the photon need not be a 
(spin unequal to zero) resonance;
in fact final states such as $\KS \pi^0 \gamma$ (for the $b \to s\gamma$ case) 
and $\pi^+ \pi^- \gamma$ (for $b \to d\gamma$) are also useable. 
In addition to making the experimental search appreciably easier 
(than if one was restricted to look for a resonance), 
this extension has other important consequences.
As stressed in~\cite{aghs}, if the signal is due to NP,
then it would be expected that the dominant contribution would be
due to the lowest order dimension 5 operator.
Then the time-dependent CP asymmetry for these three-body final states
would be independent of the energy of the photon and should
in fact coincide with that for the corresponding resonance,
allowing the statistical uncertainty to be reduced.
If, on the other hand, higher order QCD corrections are important,
then the presence of the extra gluon will cause the asymmetry to
become dependent on kinematic variables (such as the photon energy),
as well as on the specific final state.
Thus, there is a data driven method to study the effect of higher
order corrections and, in principle, their contribution
can be separated from that of the LO $H_{\rm eff}$, though of course,
the method places higher demand on luminosity.

Let us briefly mention that several other methods have been
proposed to test the polarization of the photon in $b$ decays.
In~\cite{gronauPirjolLee,GGPR} the photon polarization is probed by
considering the interferences of various $K\pi\pi$ resonances.  
The photon polarization may also be studied in $\Lambda_b$
decays~\cite{HiKa,GGPR}.
Another approach suggested in~\cite{GrossmanPirjol,Sehgal} is to
``resolve'' the photon to an $e^+e^-$ pair.
It is not yet been feasible to apply any of these methods with existing data,
which provides additional motivation to find new techniques.

One shortcoming of the oscillation signal as a probe for new physics is
that it only provides one observable. 
In fact, in principle, there exist a number of
observables related to the polarization of the photon which should be
measured and our goal here is to show how a more complete set of
observables may be obtained. 
In this paper we consider the case $B\to PV\gamma$, 
where the new feature is that the angular distribution of the
final state particles can also be used.
Under the assumption, likely true in most NP models,
that the decay is dominated by the leading order effective Hamiltonian, 
we will show that all the possible observables 
related to the photon polarization may be extracted,
subject to a 4-fold discrete ambiguity.

Among the new observables which may be obtained in this way, the
parity-odd, $T_N$-odd~\cite{tn}, triple correlations, 
resulting from the interference of left and right helicities, 
leading to {\it direct} CP asymmetries are a 
particularly powerful {\it null} test in $b \to s$ transitions.
The point is that in the SM these left and right helicities
arise predominantly from the penguin graph and
therein the CP-odd phase from the CKM-paradigm is suppressed
and is ${\cal O}(\lambda^2)$.
This suppression is in addition to the suppression of
the mixing of left and right helicities,
which according to the LO $H_{\rm eff}$ is ${\cal O}(m_s/m_b)$ 
and due to higher order QCD corrections may, 
at worst, be ${\cal O}(10\%)$~\cite{grinstein2,matsumori,bz}.
Therefore, in the SM this class of direct CP asymmetries will suffer from this
double suppression and must be $\lsim \lambda^2 \times 10\%$,
{\it i.e.} $\lsim 1\%$, giving us a class of
``golden'' observables that have negligible SM ``pollution''.

Furthermore, from the angular distributions of the $PV\gamma$ final state, 
one can also construct CP conserving observables involving
mixing of left and right photon helicities. 
These observables are likely to have SM pollution of a few percent,
so are a bit less clean but they serve a very important use 
in searching for CP-conserving NP that generates right-handed 
currents~\cite{lrmodels, lrsusy, aps}.

It should be noted that for the case of direct CP in $b \to d \gamma$
in the SM the ${\cal O}(\lambda^2)$ suppression that characterises
$b \to s \gamma$ is no longer there. 
Thus, for this case, CP-violating as well as 
CP-conserving (forward-backward) asymmetries
would typically have SM ``pollution'' of a few percent.

It is worthwhile to note more generally that the observables sensitive to
new physics in radiative decays fall broadly into two categories.
First, there are those where a large Standard Model phase 
is suppressed due to the left-handedness of the photon.
An example of this type could be the time-dependent asymmetry in 
$B_d \to \KS\pi^0\gamma$~\cite{aghs}.
While observables of this type suffer 
some pollution 
from Standard Model backgrounds,
they can play an essential role in the interpretation of new physics signals,
since they do not require the NP to carry a CP violation phase.
The second category of observables is those wherein a phase that is small
in the SM is further helicity suppressed --- these are consequently 
expected to be vanishingly small in the SM.
However, for NP to be detected, 
through these observables
it must not only possess a different current structure to the SM,
but also a new CP violation phase.
In addition to the triple correlation asymmetries discussed in this paper,
another notable observable in this category is the 
time-dependent asymmetry in $B_s \to \phi\gamma$,  
along with the corresponding non-resonant cases, such as $B_s \to K_S K_S
(K^+ K^-, \eta' \pi^0...) \gamma$~\cite{ags97,aghs}.

A particular $PV\gamma$ final state that we will focus on is $K \phi \gamma$
as it is very distinctive with many desirable features.  
In the case of charged $B$ mesons, with the subsequent decay 
$\phi\to K^+K^-$, the final state is a photon together 
with only charged kaons and so is particularly easy to detect. 
In the neutral case there is the further advantage that the oscillation
signal of~\cite{ags97,aghs} should be easier to extract.  
In particular, the $\phi$ decays right at the $B$ decay vertex,
so the vertex position is experimentally more accessible 
than for the case $B\to K^ *\gamma$, followed by $K^* \to \KS \pi^0$.
Another helpful feature of this mode is that the rate
has been measured to be relatively large 
(${\cal B}(B^\pm \to K^\pm \phi \gamma) = 
3.5 \times 10^{-6}$~\cite{phikgamma,hfag}).

In passing, we mention that a theoretical motivation for the study of 
these $PV\gamma$ final states is that 
there have been some hints~\cite{deltaSref} 
in the measurement of $\sin2\beta$ (also known as $\sin2\phi_1$)
that there could be a difference between the oscillation in
charmonium final states ({\it e.g.} $B\to \psi \KS$) 
and pure penguin states ({\it e.g.} $B\to \eta^\prime \KS$). 
If true this could mean that there is NP in the gluonic $b\to s$ penguin. 
It would therefore make sense to look for NP in
the corresponding $b\to s$ photonic penguin.
Correlations between these observables may be useful to 
elucidate the nature of the NP effect.


%
\begin{table}[htbp]
  \caption{
    $B^+$, $B^0$, $B_s \to P V \gamma$ modes where analysis of this
    paper may be applicable.
    Modes that are underlined may be particularly attractive,
    since they would be measured using final states consisting only of
    charged mesons and an energetic photon.
  }
  \begin{center}
    \begin{tabular}{c|c}
      \hline
      $b \to s\gamma$ & $b \to d\gamma$ \\

      \hline

      $B^+ \to$ \underline{$K^+ \rho^0 \gamma$}, $K^+ \omega \gamma$, \underline{$K^+ \phi \gamma$}
      & 
      $B^+ \to$ \underline{$\pi^+ \rho^0 \gamma$}, $\pi^+ \omega \gamma$, \underline{$\pi^+ \phi \gamma$}
      \\
      $B^+ \to K^0 \rho^+ \gamma$   
      & 
      $B^+ \to$ \underline{$K^+ \bar{K}^{*0} \gamma$ }
      \\
      $B^+ \to$ \underline{$\pi^+ K^{*0} \gamma$}
      &
      $B^+ \to$ \underline{$ \bar{K}^0 K^{*+} \gamma$}
      \\

      $B^+ \to \pi^0 K^{*+} \gamma$, $\eta K^{*+} \gamma$, $\eta^\prime K^{*+} \gamma$ 
      & 
      $B^+ \to \pi^0 \rho^+ \gamma$, $\eta \rho^+ \gamma$, $\eta^\prime \rho^+ \gamma$ 
      \\

      \hline

      $B^0 \to K^+ \rho^- \gamma$ 
      & 
      $B^0 \to \pi^+ \rho^- \gamma$, \underline{$K^+ K^{*-} \gamma$}  
      \\
      $B^0 \to$ \underline{$ K^0 \rho^0 \gamma$}, $K^0 \omega \gamma$, \underline{$K^0 \phi \gamma$}  
      & 
      $B^0 \to \pi^0 \rho^0 \gamma$, $\pi^0 \omega \gamma$, $\pi^0 \phi \gamma$
      \\

      $B^0 \to$ \underline{$ \pi^- K^{*+} \gamma$} 
      & 
      $B^0 \to \pi^- \rho^+ \gamma$, \underline{$K^- K^{*+} \gamma$}  
      \\

      $B^0 \to \pi^0 K^{*0} \gamma$, $\eta K^{*0} \gamma$, $\eta^\prime K^{*0} \gamma$ 
      &
      $B^0 \to \eta \rho^0 \gamma$, $\eta \omega \gamma$, $\eta \phi \gamma$
      \\
      &
      $B^0 \to \eta^\prime \rho^0 \gamma$, $\eta^\prime \omega \gamma$, $\eta^\prime \phi \gamma$
      \\

      \hline

      $B_s \to$ \underline {$ K^+ K^{*-} \gamma$} 
      & 
      $B_s \to$ \underline {$\pi^+ K^{*-} \gamma$}   
      \\

      $B_s \to \eta \rho^0 \gamma$, $\eta \omega \gamma$, $\eta \phi \gamma$
      & 
      $B_s \to$ \underline{$ \bar{K}^0 \rho^0 \gamma$}, $\bar{K}^0 \omega \gamma$, \underline{$\bar{K}^0 \phi \gamma$}     
      \\
      $B_s \to \eta^\prime \rho^0 \gamma$, $\eta^\prime \omega \gamma$, $\eta^\prime \phi \gamma$
      &
      \\

      $B_s \to$ \underline{$ K^- K^{*+} \gamma$}   
      &
      $B_s \to K^- \rho^+ \gamma$
      \\
      
      $B_s \to \pi^0 \phi \gamma$, $\eta \phi \gamma$, $\eta^\prime \phi \gamma$
      & 
      $B_s \to \pi^0 K^{*0} \gamma$, $\eta K^{*0} \gamma$, $\eta^\prime K^{*0} \gamma$ 
      \\
      \hline
    \end{tabular}
  \end{center}
  \label{many}
\end{table}

The same methodology may be used in any $B \to PV\gamma$ transition,
including $B_u$, $B_d$ and $B_s$ decays.
Table~\ref{many} lists many of the possible channels.
In the case of $B_u$ ({\it i.e.} $B^+$) decays,
the final state is always flavour specific,
and, as already noted, when the vector meson is neutral,
the final state is often particularly attractive 
from the experimental viewpoint.
For $B_d$ decays, some of the final states contain definite strangeness
({\it e.g.} $K^+ \rho^- \gamma$, or 
$\pi^0 K^{*0} \gamma$ with $K^{*0} \to K^+ \pi^-$)
and thus do not exhibit oscillations --- a time-dependent
analysis is not necessary for such channels.
Some other $B_d$ decays 
({\it e.g.} $ K^0 \rho^0 \gamma$, or
$\pi^- K^{*+} \gamma$ with $K^{*+} \to \KS \pi^+$),
are sensitive to the oscillation signal, 
and therefore gain from a time-dependent analysis.
The procedure for such an analysis is described in~\cite{aghs}.
However, the polarization signal can still be extracted from a 
time-integrated (flavour-untagged) analysis, albeit with reduced sensitivity.
In this sense, there is some similarity with the analysis of $B_s$ decays.
Moreover, in the $B_s$ case, the lifetime difference $\Delta \Gamma$ is not 
negligible, allowing some of the sensitivity to be recovered,
even when the $B_s$ oscillations cannot be resolved,
for example at a Super $B$ Factory.
Hence decay modes such as $B_s \to \pi^0 \phi \gamma$,
that will be difficult to study in the hadronic environment,
remain highly interesting.

\section{Photon Polarization}

In this paper we would like to discuss how to monitor the photon
polarization produced in the decay

\begin{eqnarray}
  b     & \to & \gamma s   \nonumber\\
  \OL b & \to & \gamma \OL s
\end{eqnarray}
\noindent
through the meson decays:
\begin{eqnarray}
  B^+     & \to & \gamma \phi K^+ \nonumber\\
  B^0     & \to & \gamma \phi \KS \nonumber\\
  \OL B^0 & \to & \gamma \phi \KS \nonumber\\
  B^-     & \to & \gamma \phi K^-
  \label{DecayList}
\end{eqnarray}
\noindent
where in each case we will take the $\phi$ to decay through the 
$\phi \to K^+K^-$ decay channel.
Note that this decay is chosen for convenience only;
all of our formalism applies equally well to decay of the vector meson,
as long as its polarization can be measured.


The main goal here is to look for signals of NP so we would like to
understand how the SM will contribute to these processes. 
Initially, we will make some assumptions concerning 
the SM contribution which should be accurate to the few percent level 
so as to make the analysis simpler. 
As we shall show, with this analysis the bulk of potential SM contamination 
to NP signals can be identified and we will then discuss 
how to improve the analysis so that the SM contamination to some NP signals 
will become ${\cal O}(1\%)$.

Initially we will make the following simplifying assumptions.

\begin{enumerate}

\item\label{assumption_one} 
  Isospin relates the neutral and charged $B$ decays in Eqn.~(\ref{DecayList}).
  This will be true if we assume that the effects of annihilation graphs 
  are small and that bremsstrahlung contributions can be neglected.

\item\label{assumption_two} 
  The decay of the $b$-quark is governed by the
  lowest order effective Hamiltonian (dimension 5) which is:

\begin{equation}
  H_{\rm eff} = - \sqrt{8} G_F \frac{e m_b}{16\pi^2} F_{\mu\nu}
  \left [
    F_L^q\ \OL q\sigma^{\mu\nu}\frac{1+\gamma_5}{2}b
    +
    F_R^q\ \OL q\sigma^{\mu\nu}\frac{1-\gamma_5}{2}b
  \right ] + h.c.
  \label{effective_H}
\end{equation}
\noindent 
where $q=s$ or $d$. 
Later we will discuss what happens when this assumption is relaxed.

If there is a strong phase, it can be taken into account, conveniently, 
by inserting it into this effective Hamiltonian.

\begin{equation}
  \hat H_{\rm eff} = - \sqrt{8} G_F \frac{e m_b}{16\pi^2} F_{\mu\nu}
  \left [
    F_L^q\ \OL q\sigma^{\mu\nu}\frac{1+\gamma_5}{2}b
    +
    F_R^q\ \OL q\sigma^{\mu\nu}\frac{1-\gamma_5}{2}b
    +
    \OL F_L^q\ \OL b\sigma^{\mu\nu}\frac{1-\gamma_5}{2}q
    +
    \OL F_R^q\ \OL b\sigma^{\mu\nu}\frac{1+\gamma_5}{2}q
  \right ]
  \label{effective_Hhat}
\end{equation}
\noindent 
where
\begin{eqnarray}
  F_R^* & = & \OL F_L\nonumber\\
  F_L^* & = & \OL F_R
  \label{cpt_conserved}
\end{eqnarray}
\noindent 
if there is no absorptive (strong) phase,
but not in general when such effects are present.

For most NP processes the dimension 5 effective Hamiltonian should
dominate. In the SM there may be difficult to calculate
corrections~\cite{grinstein1,grinstein2,bz,matsumori} 
that are likely to be a few percent, 
and in any case should be $\lsim {\cal O}(10\%)$.

\item\label{assumption_three} 
  The SM dominantly contributes only to the $F_L$ term. 
  Again this would be true in the leading order effective
  Hamiltonian where the penguin topology dominates to ${\cal O}(m_s/m_b)$. 
  As discussed in~\cite{grinstein1,grinstein2,bz,matsumori} 
  there may be additional ${\cal O}(\leq 10\%)$
  contributions to $F_R$ from the SM due to higher order effects.
\end{enumerate}

To start with we will discuss the case of the charged $B$ decays in
Eqn.~(\ref{DecayList}). The discussion generalizes to the case of the
neutral decays when the flavour of the $B$ meson can be tagged; more
generally in the neutral case there is oscillation~\cite{ags97,aghs} which
we will discuss briefly afterwords.

Assumption (\ref{assumption_two}) together with the fact that the strong
interaction is symmetric under P and C allows us to isolate the
polarization of the photon in the short distance (SD) $b\to s\gamma$
process from the subsequent long distance (LD)  hadronization to the final
state.

Suppose that a $B^\pm$ meson undergoes radiative decay:
\begin{eqnarray}
  B^+ \to \gamma Y \nonumber\\
  B^- \to \gamma \OL Y
\end{eqnarray}
\noindent
where $Y$ is a state of fixed particle content.
We can write the amplitude to left or right polarized photons as a
product of the appropriate term in the effective Hamiltonian and a
hadronization factor which depends on the details of $Y$. 
In particular,
\begin{eqnarray}
  \fm(B^-\to \gamma_R Y)     & = &      F_R g(Y)        \nonumber\\
  \fm(B^-\to \gamma_L Y)     & = & -    F_L g({\rm P}Y) \nonumber\\
  \fm(B^+\to \gamma_R \OL Y) & = & -\OL F_L g({\rm C}Y) \nonumber\\
  \fm(B^+\to \gamma_L \OL Y) & = &  \OL F_R g({\rm PC}Y)
  \label{generic_1}
\end{eqnarray}
\noindent
where C and P are charge conjugation and 
parity transformations on $Y$ respectively; 
we have included here a $-$ sign for the parity of the initial $B$ 
and another $-$ sign for 
the charge conjugation operation on the photon~\cite{aghs}.

In order to follow the effects of C and P on the hadronic state it is
useful to decompose $Y$ into a series of basis states which are also
eigenstates of P. Thus if a general state of $Y$ can be written as:
\begin{equation}
  |Y\ket = \sum_i a_i |Y_i>
\end{equation}
\noindent
Eqn.(\ref{generic_1}) then becomes:
\begin{eqnarray}
  \fm(B^-\to \gamma_R |    Y_i\ket) & = &      F_R            g_i \nonumber\\
  \fm(B^-\to \gamma_L |    Y_i\ket) & = & -    F_L  {\rm P}_i g_i \nonumber\\
  \fm(B^+\to \gamma_R |\OL Y_i\ket) & = & -\OL F_L            g_i \nonumber\\
  \fm(B^+\to \gamma_L |\OL Y_i\ket) & = &  \OL F_R  {\rm P}_i g_i
\end{eqnarray}
\noindent
where ${\rm P}_i$ is the parity eigenvalue of $|Y_i\ket$.

The generalization to neutral $B$ decays is trivial in the 
case that $Y$ is flavour specific,
and in the case that $Y$ does not distinguish between $B$ and $\OL B$ is
\begin{eqnarray}
  \fm(\OL B^0\to \gamma_R |    Y_i\ket) & = &      F_R                      g_i \nonumber\\
  \fm(\OL B^0\to \gamma_L |    Y_i\ket) & = & -    F_L  {\rm P}_i           g_i \nonumber\\
  \fm(    B^0\to \gamma_R |\OL Y_i\ket) & = & -\OL F_L  {\rm C}_i           g_i \nonumber\\
  \fm(    B^0\to \gamma_L |\OL Y_i\ket) & = &  \OL F_R  {\rm C}_i {\rm P}_i g_i
  \label{gen_amplitudes}
\end{eqnarray}
\noindent
where ${\rm C}_i$ is the charge conjugation eigenvalue of $|Y_i\ket$.

In general there are seven parameters which can be measured in $b\to s\gamma$.
These can be enumerated by noting that there are four
complex polarization amplitudes (left and right for $B$ and $\OL B$
decay) giving eight parameters from which we must subtract an overall
common unobservable phase.

It is useful, to denote these seven parameters in the following way:

\begin{eqnarray}
  {\bf F}        & = & |F_L|^2+|F_R|^2+|\OL F_L|^2+|\OL F_R|^2 \nonumber\\
  A_{CP}         & = & \frac{
    |F_L|^2+|F_R|^2-|\OL F_L|^2-|\OL F_R|^2}{
    |F_L|^2+|F_R|^2+|\OL F_L|^2+|\OL F_R|^2} \nonumber\\
  S_0            & = & \frac12 \frac{
    \Im(F_L \OL F_R^*+F_R \OL F_L^*)}{
    |F_L|^2+|F_R|^2+|\OL F_L|^2+|\OL F_R|^2} \nonumber\\
  A_{RL}         & = & \frac{
    |F_R|^2-|F_L|^2}{
    |F_R|^2+|F_L|^2} \nonumber\\
  \OL A_{RL}     & = & \frac{
    |\OL F_L|^2-|\OL F_R|^2}{
    |\OL F_L|^2+|\OL F_R|^2} \nonumber\\
  \zeta_{RL}     & = & {\rm arg}(F_R F_L^*) \nonumber\\
  \OL \zeta_{RL} & = & {\rm arg}(\OL F_R^* \OL F_L)
\end{eqnarray}

\noindent
Determining ${\bf F}$ from an exclusive $B$ decay is not practical without
a reliable model for hadronization. 
It is best measured by the inclusive $b\to s\gamma$ rate. 
Moreover, in principle $S_0$ can only be measured in the case
of neutral $B$ meson decay through oscillation methods~\cite{ags97,aghs}; 
in this case it makes sense to use instead the quantity:

\begin{equation}
  S_\phi = \frac12 \frac{
    \Im((F_L \OL F_R^*+F_R \OL F_L^*)e^{2i\phi})}{
    (|F_L|^2+|F_R|^2+|\OL F_L|^2+|\OL F_R|^2)}
\end{equation}
\noindent
where $\phi$ is the mixing angle. 
Using this notation $S_\phi$ is proportional to the quantity actually measured 
and the mixing angle 
(in the case of $B_d$ oscillations, $\phi=\beta\equiv\phi_1$) 
is presumably known from previous measurements~\cite{sin2beta}. 
If assumption~\ref{assumption_three} is correct,
$S_0$ thus deduced from measurements of neutral $B$ oscillations 
applies also to the case of charged $B$ decays.

Let us now consider specifically the case where $Y=\phi K$.
Since $Y$ is from $B\to Y\gamma$, its total angular momentum, $J$, 
can be any integer $\geq 1$. 
For each value of $J$
there are three states characterized by the helicity of the $\phi$ in the
$K\phi$ rest frame, thus the basis states are:

\begin{equation}
  |J h_\phi=+1\ket; \,\,\,  |J h_\phi=0\ket; \,\,\,   |J h_\phi=-1\ket;\ \, .
\end{equation}
\noindent
Using a basis which consists of parity eigenstates we obtain:
\begin{eqnarray}
  |J_a\ket & = & \frac{1}{\sqrt{2}}
  \left ( |J h_\phi=+1\ket- |J h_\phi=-1\ket \right) \nonumber\\
  |J_t\ket & = & \frac{1}{\sqrt{2}}
  \left ( |J h_\phi=+1\ket+ |J h_\phi=-1\ket \right) \nonumber\\
  |J_\ell\ket & = & |J h_\phi=0\ket
\end{eqnarray}
\noindent 
where the parity of these states are $(-1)^J$ for $|J_a\ket$
and $(-1)^{J+1}$ for $|J_t\ket$ and $|J_\ell\ket$

Let us denote by $p_\gamma$, $p_B$, $p_\phi$, $p_K$ the momentum of the
photon, $B$ meson, $\phi$ and $K$ respectively. 
For the subsequent decay $\phi\to K^+ K^-$,
let $q_1$ and $q_2$ be the 4-momenta of the $K^+$ and $K^-$ respectively.

To define the observed decay distributions of the final state particles,
let us introduce the following angles.
In the $K\phi$ rest frame, let us define $\eta$ to be the angle between
the momentum of the photon and the momentum of the $K$. Thus:

\begin{equation}
  \cos\eta =
  \left[
    \frac{
      \vec p_K \cdot \vec p_\gamma}{
      |\vec p_K|\ |\vec p_\gamma|}
  \right]_{K\phi\ \rm frame}
\end{equation}
\noindent 
In the $\phi$ rest frame let $\theta$ be the angle between the
momenta of the $K$ resulting from the $\phi$ decay and the $K$ produced
with the $\phi$. Thus:

\begin{equation}
  \cos\theta =
  \left[ 
    \frac{\vec p_K \cdot \vec q_1}{|\vec p_K|\ |\vec q_1|} 
  \right]_{\phi\ \rm frame}
\end{equation}
\noindent
Let $\Phi$
be the azimuthal angle of the $K^+$ with respect to the plane defined by
the $\gamma$ and $B$ momenta with the sign convention that $\sin\Phi$ is
positive when $(\vec p_\gamma \times\vec p_B)\cdot p_{K^+}$ is positive.
Thus:

\begin{equation}
  \Phi = {\rm arg}
  \left[
    {\rm Tr}(\slp_B \slp_\gamma \slp_K \slq_1 (1+\gamma_5))
  \right]
\end{equation}
\noindent
Let $s_{K\phi}$ be the invariant mass of the $K\phi$ system:
$s_{K\phi}=(p_K+p_\phi)^2$.

The amplitude for $B^-\to \gamma_R K\phi$ must be proportional to
$F_R$ so let the amplitude for $B^-\to \gamma_R |i\ket$ be $F_R g_i$.
It follows then that

\begin{eqnarray}
  \fm(B^-\to \gamma_R | i \ket) & = &      F_R           g_i \nonumber\\
  \fm(B^-\to \gamma_L | i \ket) & = & -    F_L {\rm P}_i g_i \nonumber\\
  \fm(B^+\to \gamma_R | i \ket) & = & -\OL F_L           g_i \nonumber\\
  \fm(B^+\to \gamma_L | i \ket) & = &  \OL F_R {\rm P}_i g_i \nonumber\\
\end{eqnarray}
\noindent
where ${\rm P}_i$ is the parity of the state $|i\ket$.

The angular distribution is therefore:

\begin{equation}
  \begin{array}{lcl@{\hspace{28mm}}c@{\hspace{28mm}}r}
    \fm(B^-\to\gamma K^- \phi) & = &
    \multicolumn{3}{l}{
      \frac{1}{\sqrt{2}}
      \sum_{J} g_{Ja} \Big\{ F_R
      (
      d^J_{+1,+1}(\eta)Y^{+1}_1(\theta,\phi) - 
      d^J_{-1,+1}(\eta)Y^{-1}_1(\theta,\phi)
      )
    } \\
    & & & 
    \multicolumn{2}{r}{
      + (-1)^J F_L
      (
      d^J_{+1,-1}(\eta)Y^{+1}_1(\theta,\phi) - 
      d^J_{-1,-1}(\eta)Y^{-1}_1(\theta,\phi)
      )
      \Big\}
    } \\
    & + &
    \multicolumn{3}{l}{
      \frac{1}{\sqrt{2}}
      \sum_{J} g_{Jt} \Big\{ F_R
      (
      d^J_{+1,+1}(\eta)Y^{+1}_1(\theta,\phi) + 
      d^J_{-1,+1}(\eta)Y^{-1}_1(\theta,\phi)
      )
    }\\
    & & & 
    \multicolumn{2}{r}{
      - (-1)^J F_L
      (
      d^J_{+1,-1}(\eta)Y^{+1}_1(\theta,\phi) + 
      d^J_{-1,-1}(\eta)Y^{-1}_1(\theta,\phi)
      )
      \Big\}
    } \\
    & + &
    \multicolumn{3}{l}{
      \sum_{J} g_{J\ell} \Big\{ F_R
      d^J_{0,+1}(\eta)Y^{0}_1(\theta,\phi)
      - (-1)^J
      F_l
      d^J_{0,-1}(\eta)Y^{0}_1(\theta,\phi)
      \Big\} \, .
    } \\
  \end{array}
  \label{master_eqn}
\end{equation}

For the charged $B$ decays, if you assume that the $\phi K$ state is
dominated by a finite number of partial waves then the angular
distribution will generally contain enough information to determine
$A_{CP}$, $A_{RL}$, $\OL A_{RL}$, $\zeta_{RL}$ and $\OL \zeta_{RL}$.
Oscillations in neutral $B$ mesons are needed to determine $S_0$ if
assumption~\ref{assumption_one} is true.

\section{\boldmath Toy Model: $J=1$}

As a toy model to illustrate this, let us make the {\it ad hoc} assumption 
that the $\phi \KS$ system is saturated by $J=1$ states. 
Thus the system is assumed to be a linear combination of:

\begin{equation}
  \{ |1_a\ket, \,\,\, |1_t\ket, \,\,\, |1_\ell\ket \} \, .
\end{equation}

If we expand Eqn.~(\ref{master_eqn}) for this case and square it to
determine the angular distribution, we obtain:


%
%

\begin{eqnarray}
  \frac{1}{\Gamma}
  \frac{d\Gamma(B^-\to K^- \phi \gamma)}
  {d\cos\eta \, d\cos\theta \, d\Phi} 
  & = &
  \frac{16\pi}{3}
  \big(
  \lambda_0 \sin^2\theta
  + \lambda_1 \sin^2\theta\cos^2\eta
  + \lambda_2 \sin^2\theta\cos\eta
  \nonumber\\
  & + & 
  \lambda_3 \sin^2\theta\cos(2\Phi)
  + \lambda_4 \sin^2\theta\sin(2\Phi)
  + \lambda_5 \sin^2\theta\cos\eta\cos(2\Phi)
  \nonumber\\
  & + &
  \lambda_6 \sin^2\theta\cos\eta\sin(2\Phi)
  + \lambda_7 \sin^2\theta\cos^2\eta\cos(2\Phi)
  + \lambda_8 \sin^2\theta\cos^2\eta\sin(2\Phi)
  \nonumber\\
  & + &
  \lambda_9\cos^2\theta\sin^2\eta
  + \lambda_{10} \sin 2\theta\sin\eta\cos\Phi
  + \lambda_{11} \sin 2\theta\sin\eta\sin\Phi
  \nonumber\\
  & + & 
  \lambda_{12} \sin2\theta\sin2\eta\cos\Phi
  + \lambda_{13}   \sin2\theta\sin2\eta\sin\Phi
  \big)
  \label{ang_dist}    
\end{eqnarray}

\noindent
where $\lambda_0$, $\lambda_1$ and $\lambda_6$ are subject to the
normalization condition:

\begin{equation}
  \lambda_0+\frac13 \lambda_1+\frac13 \lambda_6=1
\end{equation}

%
%

In terms of the $F$ and $g$ couplings, the coefficients of the above
angular distribution are:

\begin{eqnarray}
  \lambda_0 & = &
  \frac12\left(|g_{1t}|^2|F_R+F_L|^2+|g_{1a}|^2|F_R-F_L|^2 \right)
  \nonumber\\
  \lambda_1 & = &
  \frac12\left(|g_{1t}|^2|F_R-F_L|^2+|g_{1a}|^2|F_R+F_L|^2 \right)
  \nonumber\\
  \lambda_2 & = &
  2\,\Re(g_{1a}g_{1t}^*)(|F_R|^2+|F_L|^2)
  \nonumber\\
  \lambda_3 & = &
  \frac12\left(-|g_{1t}|^2|F_R+F_L|^2+|g_{1a}|^2|F_R-F_L|^2 \right)
  \nonumber\\
  \lambda_4 & = &
  \Im(g_{1a}g_{1t}*)(|F_R|^2-|F_L|^2)+2\,\Re(g_{1a}g_{1t}*)\Im(F_RF_L^*)
  \nonumber\\
  \lambda_5 & = &
  -4\,\Re(g_{1a}g_{1t}*)\Re(F_RF_L^*)
  \nonumber\\
  \lambda_6 & = &
  2 (|g_{1t}|^2+|g_{1a}|^2) \Im(F_RF_L^*)
  \nonumber\\
  \lambda_7 & = &
  \frac12\left(|g_{1t}|^2|F_R-F_L|^2-|g_{1a}|^2|F_R+F_L|^2 \right)
  \nonumber\\
  \lambda_8 & = &
  -\Im(g_{1a}g_{1t}^*)(|F_R|^2-|F_L|^2)+2\,\Re(g_{1a}g_{1t}*)\Im(F_RF_L^*)
  \nonumber\\
  \lambda_9 & = &
  |g_{1\ell}|^2|F_R-F_L|^2
  \nonumber\\
  \lambda_{10} & = &
  -\Re(g_{1a}g_{1\ell}^*)|F_R-F_L|^2
  \nonumber\\
  \lambda_{11} & = &
  \Im(g_{1t}g_{1\ell}^*)(|F_R|^2-|F_L|^2)-2\,\Re(g_{1t}g_{1\ell}^*)\Im(F_RF_L^*)
  \nonumber\\
  \lambda_{12} & = &
  -\frac12 \,\Re(g_{1t}g_{1\ell}^*)|F_R-F_L|^2
  \nonumber\\
  \lambda_{13} & = &
  \frac12 \,\Im(g_{1a}g_{1\ell}^*)(|F_R|^2-|F_L|^2)-\Re(g_{1a}g_{1\ell}^*)\Im(F_RF_L^*)
  \label{lambda_def}
\end{eqnarray}

%
%

The quantities $A_{RL}$ and $\zeta_{RL}$ may be extracted from these
coefficients in a number of ways. In particular, let us focus on
$\lambda_{0-8}$. These 9 quantities are all functions of the four complex
numbers $\{g_{1a}$, $g_{1t}$, $F_R$, $F_L\}$ and therefore eight real
numbers. None of the observables $\lambda_{0-8}$ are altered by the
transformation in terms of three arbitrary real constants $\{A,\ B,\ C\}$:
\begin{eqnarray}
g_{1a} &\to& e^{A+iB}g_{1a}\nonumber\\
g_{1t} &\to& e^{A+iB}g_{1t}\nonumber\\
F_R &\to& e^{-A+iC}F_R\nonumber\\
F_L &\to& e^{-A+iC}F_L
\end{eqnarray}
\noindent
so there are therefore only a net of 4 real unknowns; besides, this
implies that there are four constraints between these observables. 
These can be written:
\begin{eqnarray}
\lambda_0^2-\lambda_1^2 &=& \lambda_3^2-\lambda_7^2
\nonumber\\
(\lambda_4+\lambda_8)(\lambda_0+\lambda_1) &=& \lambda_2\lambda_6
\nonumber\\
\lambda_5(\lambda_0+\lambda_1) &=& \lambda_2(\lambda_7+\lambda_3)
\nonumber\\
(\lambda_4-\lambda_8)^2(\lambda_0-\lambda_1)^2
&=&
\left [
(\lambda_0+\lambda_1)^2-(\lambda_7-\lambda_3)^2-\lambda_2^2
\right ]
\left [
(\lambda_0+\lambda_1)^2-(\lambda_7+\lambda_3)^2-\lambda_6^2
\right ]
\label{dependency}
\end{eqnarray}

Thus there are 5 linearly independent observables which should
allow us to determine the 4 unknowns.

Extracting the quantities $A_{RL}$ and $\zeta_{RL}$ we obtain:
\begin{eqnarray}
  \frac12\sqrt{1-A_{RL}^2} \cos\zeta_{RL}  & = & \Re(Q_{RL})
%
  = \frac12 \left( \frac {\lambda_0-\lambda_1} {\lambda_7-\lambda_3} \right)
  \label{eqna}\\
  \frac12\sqrt{1-A_{RL}^2} \sin\zeta_{RL}  & = & \Im(Q_{RL})
  = \frac12 \left( \frac {\lambda_6}{\lambda_0+\lambda_1} \right)
  \label{eqnb}\\
  A_{RL}^2 & = & \left( \frac{|F_R|^2-|F_L|^2}{|F_R|^2+|F_L|^2} \right)^2
  = \frac {\lambda_4-\lambda_8}{
    \sqrt{(\lambda_0+\lambda_1)^2-(\lambda_7-\lambda_3)^2-\lambda_2^2}}
  \label{eqnc}
\end{eqnarray}
\noindent
where,
\begin{eqnarray}
  Q_{RL} & = & \frac{F_R F_L^*}{|F_R|^2+|F_L|^2}
  \label{qrl}
\end{eqnarray}

\noindent
Since only the square
of $A_{RL}$ is fixed by the data, the sign of $A_{RL}$ is not fixed by the
data giving a two fold ambiguity in the polarization parameters.

Let us mention in passing that actually there are additional
5 observables, {\it i.e.} $\lambda_9$ to $\lambda_{13}$ which
involve 2 more unknowns due to $g_{1l}$; thus there is
considerable redundancy which should help in solving for
all the unknowns :  $\{g_{1a}$, $g_{1t}$, $g_{1l}$, $F_R$, $F_L\}$

Once $F_{R,L}$ have been determined, then $\Re(Q_{RL})$ (see Eqn.\ref{qrl})
is only allowed to be a few percent in the SM 
so that should enable us to directly test the SM.
This test does not involve CP; it only allows determination
of the photon polarization. Amongst the 14 observables listed
in Eqn.~\ref{lambda_def} there are several that directly
allow determination of $\Re\,F_R$, see {\it e.g.} $\lambda_5$
which monitors a forward backward asymmetry in the angular
distribution, eq.~\ref{ang_dist}.

Turning now to $\Im(Q_{RL})$, the SM can only accommodate it to be
$\lsim 1\%$ provided it is genuinely CP-violating. 
Since it results
in the above list of observables from triple correlation asymmetries,
these are not necessarily CP-violating. To address this issue,
it is important to compare $B^-$ and $B^+$ data in order to distinguish
the effect of strong phases from truly CP violating phases. 
In our toy model, the angular distribution 
in the case of $B^+$ decay is given in the
same form with $\lambda_i$ replaced by $\OL \lambda_i$.
The coefficients of this distribution are given in terms of the
couplings as in Eqn.~(\ref{lambda_def}).

\begin{eqnarray}
  \OL \lambda_0 & = &
  \frac12\left(|g_{1t}|^2|\OL F_R+\OL F_L|^2+|g_{1a}|^2|\OL F_R-\OL F_L|^2 \right)
  \nonumber\\
  \OL \lambda_1 & = &
  \frac12\left(|g_{1t}|^2|\OL F_R-\OL F_L|^2+|g_{1a}|^2|\OL F_R+\OL F_L|^2 \right)
  \nonumber\\
  \OL \lambda_2 & = &
  2\,\Re(g_{1a}g_{1t}^*)(|\OL F_R|^2+|\OL F_L|^2)
  \nonumber\\
  \OL \lambda_3 & = &
  \frac12\left(-|g_{1t}|^2|\OL F_R+\OL F_L|^2
    +|g_{1a}|^2|\OL F_R-\OL F_L|^2 \right)
  \nonumber\\
  \OL \lambda_4 & = &
  \Im(g_{1a}g_{1t}^*)(|\OL F_R|^2-|\OL F_L|^2)
  +2\,\Re(g_{1a}g_{1t}*)\Im(\OL F_R\OL F_L^*)
  \nonumber\\
  \OL \lambda_5 & = &
  -4\,\Re(g_{1a}g_{1t}*)\Re(\OL F_R\OL F_L^*)
  \nonumber\\
  \OL \lambda_6 & = &
  2 (|g_{1t}|^2+|g_{1a}|^2) \Im(\OL F_R \OL F_L^*)
  \nonumber\\
  \OL \lambda_7 & = &
  \frac12\left(|g_{1t}|^2|\OL F_R-\OL F_L|^2
    -|g_{1a}|^2|\OL F_R+\OL F_L|^2 \right)
  \nonumber\\
  \OL \lambda_8 & = &
  -\Im(g_{1a}g_{1t}*)(|\OL F_R|^2-|\OL F_L|^2)
  +2\,\Re(g_{1a}g_{1t}*)\Im(\OL F_R\OL F_L^*)
  \nonumber\\
  \OL \lambda_9 & = &
  |g_{1\ell}|^2|\OL F_R-\OL F_L|^2
  \nonumber\\
  \OL \lambda_{10} & = &
  -\Re(g_{1a}g_{1\ell}^*)|\OL F_R-\OL F_L|^2
  \nonumber\\
  \OL \lambda_{11} & = &
  \Im(g_{1t}g_{1\ell}^*)(|\OL F_R|^2-|\OL F_L|^2)
  -2\,\Re(g_{1t}g_{1\ell}^*)\Im(\OL F_R\OL F_L^*)
  \nonumber\\
  \OL \lambda_{12} & = &
  -\frac12 \, \Re(g_{1t}g_{1\ell}^*)|\OL F_R-\OL F_L|^2
  \nonumber\\
  \OL \lambda_{13} & = &
  \frac12 \, \Im(g_{1a}g_{1\ell}^*)(|\OL F_R|^2-|\OL F_L|^2)
  -\Re(g_{1a}g_{1\ell}^*)\Im(\OL F_R\OL F_L^*)
  \label{lambda_def_bar}
\end{eqnarray}

From this we can find $\OL A_{RL}$ and $\OL \zeta_{RL}$
as in Eqns.~(\ref{eqna}, \ref{eqnb}, \ref{eqnc}) where we also have
a two fold
ambiguity in the sign of $\OL A_{RL}$.

Let us consider the relation between the $B^-$ and $B^+$ decay in the case
where there is no strong phase in $F_L$ or $F_R$ so that
Eqn.~(\ref{cpt_conserved})
applies. In this case the coefficients of the
P-even terms must also be C-even, thus

\begin{eqnarray}
\lambda_{0,1,2,3,5,7,9,10,12}
=\OL \lambda_{0,1,2,3,5,7,9,10,12}
\label{parityeven}
\end{eqnarray}

\noindent 
The P-odd terms, in general, have two components, either C-even or C-odd.  
The P-odd C-even parts are CP violating and so proportional to
the sine of the weak phase between $F_R$ and $F_L$, for instance, the
second in the expression for $\lambda_4$. 
The C-odd, and therefore CP-even component is proportional 
to the sine of the strong phase difference between the $g_i$ terms. 
In this way, the quantities:

\begin{eqnarray}
\lambda_4+\OL \lambda_4,\
\lambda_6+\OL \lambda_6,\
\lambda_8+\OL \lambda_8,\
\lambda_{11}+\OL \lambda_{11},\
\lambda_{13}+\OL \lambda_{13}
\end{eqnarray}

\noindent
are CP violating and proportional to $\sin\zeta_{RL}$ while

\begin{eqnarray}
\lambda_4-\OL \lambda_4,\
\lambda_6-\OL \lambda_6,\
\lambda_8-\OL \lambda_8,\
\lambda_{11}-\OL \lambda_{11},\
\lambda_{13}-\OL \lambda_{13}
\end{eqnarray}

\noindent
are C-odd P-odd and so proportional to $A_{RL}$.

Even if there are strong phases in $F_R$ or $F_L$ and so one does not have
Eqn.~(\ref{cpt_conserved}), in Eqns.~(\ref{eqna}, \ref{eqnb}, \ref{eqnc}) we
showed that one could extract $\sin\zeta_{RL}$ and $\sin\OL \zeta_{RL}$
separately.  This information is particularly useful in detecting the
presence of NP and distinguishing it from the SM contribution.  As
already mentioned many times,
by assumption \ref{assumption_two}, right polarized photon in $b$ decay (and
left in $\OL b$ decay) is suppressed in the SM.  Thus signals which are
proportional to this polarization may suggest the presence of NP if
they are larger than the small background in the SM expected to be
less than a few percent~\cite{grinstein1,grinstein2,bz,matsumori}.
However, the main mechanism for producing SM ``pollution''
in~\cite{grinstein1, grinstein2} would produce photons with
the same CP phase as the dominant SM photons up to corrections
of ${\cal O}(\lambda^2)$, {\it i.e.} $\approx  5\%$.
This would mean that the ratio
$\sin\zeta_{RL}$ in Eqn.~(\ref{eqnb}) would be further suppressed by 
$\sim {\cal O}(\lambda^2)$ 
and so now the SM background to the CP violating signal due
to NP is $\lsim 1\%$.
Note that, in principle, $\sin\zeta_{RL}$ and $\sin\OL \zeta_{RL}$
could receive contributions from strong phases. 
The combination $\sin\zeta_{RL}-\sin\OL \zeta_{RL}$
is CP violating and so would be immune from contributions of this type.

\section{Saturation of Partial Waves}

In order to use a finite number of partial waves such as the toy model
used above, it is helpful to have a measure of whether the number of
partial waves used adequately describes the data. 
Referring to Eqn.~(\ref{master_eqn}) we see that if partial waves up to angular
momentum $J_{\rm max}$ are used then $d\Gamma/d \cos\eta$ will be a polynomial
in $\cos\eta$ of degree $2 J_{\rm max}$. 
We can test whether a data set can be described in this way 
by constructing a kernel for such distributions. 
If $P^J(z)$ is the Legendre polynomial of degree $J$, then let us define:

\begin{equation}
  K_J(x,y) = \sum_{m=0}^{J_{\rm max}} \frac{2}{4m+1}P^{2m}(x)P^{2m}(y) \, .
\end{equation}
\noindent
The first few such kernels are:

\begin{eqnarray}
  K_0(x,y) & = & \frac12   \nonumber\\
  K_1(x,y) & = & \frac38 \left[ 3+15x^2y^2-5(x^2+y^2) \right] \nonumber\\
  K_2(x,y) & = & \frac{15}{128} \left[ 
    15 - 70(x^2+y^2) + 588x^2y^2 + 63(x^4+y^4)+735x^4y^4 - 630x^2y^2(x^2+y^2)
  \right] \, .
\end{eqnarray}

If $f(x)$ is a polynomial of degree $\leq 2J$ then
\begin{eqnarray}
  \int_{-1}^{+1}\int_{-1}^{+1} K(x,y) f(x)f(y)\,dx\,dy = \int_{-1}^{+1}f^2(x)dx \, .
\end{eqnarray}

Thus, if one measures the
value of $\eta$ in $n$ events, then if
\begin{eqnarray}
  \frac{1}{n^2}\sum_{i=1,j=1}^{i=n,j=n}
  K_J(\cos\eta_i,\cos\eta_j)
\end{eqnarray}

\noindent
is equal to 1 within experimental error then partial waves up to $J$
are adequate to explain the data.

In any case, the net effect of truncating the partial wave expansion on
the key signal for NP, $\sin\zeta_{RL}$ will be to somewhat alter the
fitted value but any P-odd C-even signal has unique symmetry
properties that tend to flag the presence of NP.

\section{\boldmath Neutral $B$ Meson}

Thus far we have considered mainly the case of charged $B$ decays
to $PV\gamma$ final states, {\it e.g.} $B^\pm\to K^\pm\phi\gamma$. 
In the case of the analogous neutral $B$ decay, there
is the added feature that the system undergoes $B\OL B$ oscillation.

In an experiment, there are three different approaches one might take to
oscillation in $B^0\to \KS\phi\gamma$:

\begin{enumerate}
\item\label{inclusive}
  For a tagged sample of $B^0$-mesons, find the time-dependent decay rate for
  $B\to \KS\phi\gamma$.
\item\label{untagged}
  Take an untagged sample and fit the angular distribution in $\eta$,
  $\theta$ and $\Phi$.
\item\label{wholehog}
  Take a tagged sample and fit the time-dependent angular distribution 
  in $\eta$, $\theta$ and $\Phi$.
\end{enumerate}

Case~\ref{inclusive} was discussed in~\cite{aghs}. 
As explained there this kind of study
allows us to extract the photon polarization 
without the need to study the angular distribution.

Let us consider case~\ref{untagged} which is identical to taking an
incoherent mixture of $B^0$ and $\OL B^0$ in the initial state. In our toy
model then, the components of the angular distribution will be:

\begin{equation}
  \tilde \lambda_i = \frac12 \left( \lambda_i+\OL \lambda_i \right)
\end{equation}

If we assume that there are no strong phases in $F_{L,R}$, since the
parity even distributions are also C-even Eqn~(\ref{parityeven}) applies,
so $\tilde\lambda_{0,1,2,3,5,7,9,10,12} =\lambda_{0,1,2,3,5,7,9,10,12}$.
For the parity odd components of the distribution, only the
C-even P-odd part survives and so in the toy model, these coefficients
are:

\begin{eqnarray}
  \lambda_4 & = &
  2\,\Re(g_{1a}g_{1t}*)\Im(F_RF_L^*)
  \nonumber\\
  \lambda_6 & = &
  2 (|g_{1t}|^2+|g_{1a}|^2)  \Im(F_RF_L^*)
  \nonumber\\
  \lambda_8 & = &
  +2\,\Re(g_{1a}g_{1t}*)\Im(F_RF_L^*)
  \nonumber\\
  \lambda_{11} & = &
  -2\,\Re(g_{1t}g_{1\ell}^*)\Im(F_RF_L^*)
  \nonumber\\
  \lambda_{13} & = &
  -\Re(g_{1a}g_{1\ell}^*)\Im(F_RF_L^*)
  \label{lambda_def_CevenPodd}
\end{eqnarray}

\noindent
The results in Eqns.~(\ref{eqna}, \ref{eqnb})  go through
unmodified since the ratios used there do not involve the C-odd, P-odd
component of the coefficients:

\begin{eqnarray}
  \frac12\sqrt{1-A_{RL}^2} \cos\zeta_{RL} & = & \Re(Q_{RL})
  = \frac12 \left( 
    \frac {\tilde\lambda_0-\tilde\lambda_1} {\tilde\lambda_7-\tilde\lambda_3}
  \right) = -\frac12\left( 
    \frac {\tilde\lambda_7+\tilde\lambda_3} {\tilde\lambda_0+\tilde\lambda_1}
  \right)
  \label{eqna_tilde}\\
  \frac12\sqrt{1-A_{RL}^2} \sin\zeta_{RL} & = & \Im(Q_{RL})
  = \frac12 \left( 
    \frac {\tilde \lambda_6}{\tilde \lambda_0+\tilde \lambda_1} 
  \right) = \frac12 \left( 
    \frac{
      (\tilde \lambda_4+\tilde\lambda_8)(\tilde\lambda_7+\tilde\lambda_3)} {
      (\tilde\lambda_0+\tilde\lambda_1)\tilde\lambda_5
    } 
  \right)
  \label{eqnb_tilde}
\end{eqnarray}

The analog to Eqn.~(\ref{eqnc}) does not work because all of the terms
proportional to $A_{LR}$ will be P-odd and C-odd and will therefore cancel
in the incoherent sum. 
We can indirectly infer it from the above results since:

\begin{eqnarray}
  A_{RL}^2 = 1 - 4\left(\frac{\Re(F_R F_L^*)}{|F_R|^2+|F_L|^2}\right)^2
  - 4\left(\frac{\Im(F_R F_L^*)}{|F_R|^2+|F_L|^2}\right)^2
\end{eqnarray}

If in fact there are significant strong phases in $F_{L,R}$, then one
needs to consider the time dependence of the angular distribution in order
to fully determine all of the photon polarizations. The time-dependent
angular distribution may be written as:
\begin{eqnarray}
  \Gamma(\OL B^0 \to \KS \phi \gamma)[t,\eta,\theta,\Phi]
  & \propto &
  \left[
    X[\eta,\theta,\Phi] + 
    Y[\eta,\theta,\Phi]\cos(\Delta m t) + 
    Z[\eta,\theta,\Phi]\sin(\Delta m t)
  \right] e^{-|t|/\tau}
  \nonumber\\
  \Gamma(    B^0 \to \KS \phi \gamma)[t,\eta,\theta,\Phi]
  & \propto &
  \left[
    X[\eta,\theta,\Phi] - 
    Y[\eta,\theta,\Phi]\cos(\Delta m t) - 
    Z[\eta,\theta,\Phi]\sin(\Delta m t)
  \right] e^{-|t|/\tau} \, ,
  \label{bd_pdfs}
\end{eqnarray}
\noindent 
where $\Delta m$ and $\tau$ are the mixing parameter 
and the neutral $B$ lifetime (here in the $B_d$ system).
Note that in the above we have not included the possible dependence 
on the invariant masss of the ($\KS\phi$) hadronic system~\cite{aghs}.
If we determine $X$ and $Y$ then $X+Y$ is the decay distribution
for $\OL B^0$ if there were no oscillations while $X-Y$ is the decay
distribution for $B^0$ without oscillation. 
The analysis thus reduces to the case for charged $B$ mesons.

%
%
%
%

\section{\boldmath Application in $B_s$ decays}

As listed in Table~\ref{many},
a large number of $B_s \to PV\gamma$ decays can potentially be used.
Similar arguments as for the $B_d$ case apply, 
but with a couple of notable differences.
Firstly, in the $B_s$ system, the lifetime difference $\Delta \Gamma$
is not negligible, and so Eqns.~\ref{bd_pdfs} has to be replaced by~\cite{aghs}

\begin{equation}
  \begin{array}{l@{\hspace{75mm}}c@{\hspace{75mm}}r}
    \multicolumn{2}{l}{
      \Gamma(\OL B^0 \to \KS \phi \gamma)[t,\eta,\theta,\Phi]
      \propto 
      \Big[
      X[\eta,\theta,\Phi]\cosh(\frac12 \Delta \Gamma t) + 
      Y[\eta,\theta,\Phi]\cos(\Delta m t)
    } \\
    & 
    \multicolumn{2}{r}{
      +
      Z[\eta,\theta,\Phi]\sin(\Delta m t) + 
      W[\eta,\theta,\Phi]\sinh(\frac12 \Delta \Gamma t)
      \Big] e^{-|t|/\tau}
    } \\
    \multicolumn{2}{l}{
      \Gamma(    B^0 \to \KS \phi \gamma)[t,\eta,\theta,\Phi]
      \propto
      \Big[
      X[\eta,\theta,\Phi]\cosh(\frac12 \Delta \Gamma t) - 
      Y[\eta,\theta,\Phi]\cos(\Delta m t)
    }\\
    &
    \multicolumn{2}{r}{
      -
      Z[\eta,\theta,\Phi]\sin(\Delta m t) + 
      W[\eta,\theta,\Phi]\sinh(\frac12 \Delta \Gamma t)
      \Big] e^{-|t|/\tau} \, .
    } \\
  \end{array}
  \label{bs_pdfs}
\end{equation}
\noindent 
Therefore in this case, an untagged and time-independent analysis
can retain some sensitivity to the photon polarization,
even without an angular analysis, through the $W$ term.
This would be particularly important at an $e^+e^-$ collider,
producing $B_s$ mesons by running at the 
$\Upsilon(5{\rm S})$ energy~\cite{y5s},
whereby it will be difficult to resolve the $\Delta m_s$ oscillations.
In this case, the observed decays are an incoherent mixture of
$B_s$ and $\OL B_s$, so that the $Y$ and $Z$ terms cancel,
and with them the sensitivity to the C-odd components.
Note, however, that the C-even P-odd component of the distribution
will not cancel. 
This is proportional to $\sin\zeta_{RL}$, 
which is the quantity most sensitive to NP.
Of course, additional sensitivity will be gained by studying the 
angular distributions.
For any of the $B_s \to PV\gamma$ decays 
that can be measured in a hadronic environment, 
the maximum sensitivity will be gained by determining also the 
$Y$ and $Z$ terms, which depend on $\cos(\Delta m t)$ and $\sin(\Delta m t)$
respectively.

\section{SM Backgrounds}

There are three classes of SM backgrounds that we need to be concerned
with:

\begin{enumerate}

\item\label{brems}
  Radiation from $b\to s$ strong penguin processes.

\item\label{tree}
  Radiation from tree $b\to u s \OL u$ processes which therefore have a
  different CKM phase.

\item\label{annih}
  Photons produced from a graph with annihilation topology.

\end{enumerate}

Backgrounds of type~\ref{brems} were discussed 
in~\cite{grinstein1,grinstein2,bz,matsumori} in the context of
the $K^* \gamma$ final state. 
As already mentioned, these estimates vary considerably from
negligible~\cite{matsumori,bz} to 10\%~\cite{grinstein2}.
Such higher order corrections, which we label as $\delta_{\rm hoc}$
would be expected to contribute to right polarized photons in
$b$-quark decays so there will be photons of the opposite helicity
contradicting assumption~\ref{assumption_one}. 
Furthermore this mechanism is not described by the effective 
Lagrangian Eqs.~(\ref{effective_H}, \ref{effective_Hhat}) so assumption
\ref{assumption_two} will not be true either.
Although there are significant uncertainties in estimating the size
of these corrections, in $b \to s$ transitions, the photons from
this source will, however, predominantly have the same CP-odd CKM
phase as the SM photonic penguin {\it i.e.} ${\cal O}(\lambda^2)$. 
Thus, CP violating triple correlation asymmetries generated from this
background suffer from two suppression factors: $\delta_{\rm hoc}$,
which reflects the suppressed photon helicity and the other is the
CP-odd phase ${\cal O}(\lambda^2)$ and are therefore expected to be of
${\cal O}(\delta_{\rm hoc} \times \lambda^2) \lsim 1\%$, 
therefore these lead to a class of excellent null tests.


There are several ways to look for NP in spite of this background
and there are some checks available to see if this effect influences
the results.

First of all, triple correlations asymmetries relevant to our
$PV\gamma$ final states which can lead to (P-odd C-even) CP
asymmetries are monitored by $\Im(Q_{RL})$, see Eqn.~(\ref{qrl}). 
However, for such asymmetries to be recognized as genuine 
CP asymmetries a comparison of $B$ and $\bar B$ decays is mandatory. 
CP asymmetries significantly more than 1\% 
would be an unambiguous signal of NP.


Recall that these final states also allow tests of the SM via
(CP-conserving) observables that monitor the interference between
the LH and RH photons in $b$-quark decays, for example, the
forward-backward (FB) asymmetry. 
These are monitored by $\Re(Q_{RL})$.
The SM pollution for these from the above type ~\ref{brems}
background is again of ${\cal O}(\delta_{\rm hoc})$, {\it i.e.} a few percent.
The underlying reason for suppression of such interference between LH and RH
photons is the fact that weak interactions in the SM are left-handed. 
Therefore, FB asymmetries significantly bigger than
$\delta_{\rm hoc}$ would be a signal of NP. 
This test is especially relevant to sources of NP which gives rise to RH
currents such as the LRSM~\cite{lrmodels} 
or in some versions of SUSY~\cite{lrsusy}
or in warped extra-dimensions~\cite{aps}.

Note also that the background  photons of type ~\ref{brems}
constituting SM pollution violate assumption~\ref{assumption_two} and
therefore their contamination will result in a dependence of 
the extracted parameters on the kinematics of the final state
({\it i.e.} on the invariant mass of the $PV$ system 
or equivalently the energy of the photon).
In particular, both CP violating triple correlations and 
forward-backward asymmetries will depend on the
energy of photon if they originate from SM pollution, 
{\it i.e.} from higher order QCD corrections. 
On the other hand, NP is expected
to show up only via modification of the dimension 5 $H_{\rm eff}$.


Contamination of type~\ref{tree} will have a CP phase difference
with the SM photonic penguin and so are a potential contamination
even to $\Im(Q_{RL})$.
For the $b \to s$ the tree is CKM suppressed with respect to the
penguin to start with, and the usual hard energy cut on $E_{\gamma}$
should be very effective in reducing this background significantly.

Contamination of type~\ref{annih} involving annihilation are larger
for charged $B$ decays than for neutral $B$ mesons ($B_d$ and $B_s$). 
Therefore to a large extent this background can be 
avoided by examining the neutral case. 
Recall also that the annihilation photons in $b$-quark decays
are expected to predominantly have the same (LH) helicity as dominant
penguin contribution~\cite{aghs}. 
In $b$-decays the RH helicity photons from annhilations graphs are suppressed. 
The overall size of the annihilation amplitude is much less compared 
to the penguin to begin with~\cite{abs,gp00} 
and furthermore for the $b \to s$ case, there is also a CKM-suppression. 
Therefore despite the large CKM phase present in the annihilation graph, 
this background is expected to be very small.




\section{Summary \& Outlook}

Radiative $B$ decays should provide an important venue to look for
signs of new physics. 
In inclusive $B$ decays one can measure the rate 
and the overall CP asymmetry of $b\to s(d) \gamma$ and compare
those measurements to the predictions of the SM. 
Exclusive $B$ decays allow for the determination of the 
remaining 5 polarization observables that can provide additional 
valuable tests of the SM and may also be used to discriminate 
between different types of New Physics, once it is discovered.
Given also that there are often experimental advantages in dealing
with exclusive modes, these avenues are clearly very worthwhile.

In the previous work~\cite{ags97,aghs} it was shown that by the
study of oscillation in neutral $B$ to a photon and to a hadronic state
of definite charge conjugation  will give an important  signal which
is particularly sensitive to NP. 
Some examples of relevant final states are 
$K^{*0}(892)\gamma$, 
$\rho^0(770)\gamma$, $\omega\gamma$, 
$K^0 \pi^0 \gamma$, $K^0 \eta \gamma$, $K^0 \phi \gamma$, 
$\pi^+ \pi^- \gamma$ {\it etc}. 
As explained in~\cite{aghs} for cases such as 
$K^0 \phi\gamma$ and $K^0 \rho\gamma$ 
no angular analysis is needed to extract the time-dependent CP asymmetry. 
According to~\cite{grinstein1,grinstein2,bz,matsumori} 
these final states may receive contamination from SM higher order
effects of a few percent. 
Although, the precise value of these higher order contaminations 
is rather difficult to calculate, as stressed in~\cite{aghs}, 
there are data driven handles to overcome this SM pollution.

When the hadron accompanying the final photon is just a single
particle, such as $K^{*0}(890)$ 
then the lowest order $H_{\rm eff}$ predicts that 
the time-dependent CP asymmetry will be the same ($\approx 2m_s/m_b$)
for every case, independent of the mass or the $J^{PC}$ of that state. 
More strikingly, to the extent that LO $H_{\rm eff}$ is valid the
time-dependent CP asymmetry will be independent of Dalitz variables
({\it i.e.} invariant mass of the hadronic state or the energy of
the photon) and will still be $\approx 2 m_s/m_b$
 when we have a multiparticle self-conjugate hadronic final state
(such as $\KS \pi^0$, $\KS \eta$, $\KS \eta^\prime$)
accompanying the radiated photon. 
To the extent that the LO $H_{\rm eff}$
is valid the statistics of the time-dependent CP asymmetry can
therefore be significantly improved by adding the results from all
such cases mentioned above. 
On the other hand, SM higher order corrections will contribute 
differently to each resonant state depending on its mass 
and also on its $J^{PC}$.
Furthermore, for the case of multihadron final states,
this SM pollution will depend also on the energy of the photon. 
Therefore, given enough data one should be
able to determine the part of the time-dependent CP asymmetry that
stays constant versus the part that varies with the accompanying
hadron and/or the photon energy.  
Hence, in principle, there is a data driven, clean procedure 
for searching for NP through such time-dependent CP studies.

In this paper we extend these previous works to include exclusive
final states that contain a vector particle ({\it i.e.} $PV\gamma$)
allowing for a non-trivial angular analysis and the use of
(time-independent) triple correlation CP asymmetries.
We show that by suitable angular analysis in
$B\to\phi K\gamma$ and other $PV\gamma$ final states the remaining
four polarization observables may be determined. In particular, for
all such $b \to s$ transitions, the P-odd, C-even triple correlation
asymmetries will have less contamination,
($\lambda^2\delta_{\rm hoc}<{\cal O}(1\%)$), 
from the SM and hence constitute an excellent {\it null test},
or ``golden'' observable signifying a very clean probe of NP. 
The particular final state $\phi K \gamma$ has
the further advantage that it has a relatively large branching
ratio~\cite{phikgamma} and leads to a final state that
is readily detectable. 
In the neutral case, the P-odd, C-even signal does not need 
flavour tagging or time-dependent measurements. 
Incidentally, if the $\phi$ decays to $K^+K^-$ the decay vertex 
may be localized and so this final state 
is also well suited for oscillation studies.
For the case of $B^\pm \to \phi K^\pm \gamma$, there is also the
advantage that on decays of the $\phi$ there will be three charged kaons
and only one photon so this final state may well even be accessible
to hadron colliders. 

There is a plethora of channels in $B$ decays (see Table~\ref{many}) 
wherein the current analysis is applicable.
Several of these have the attractive feature of having 
only charged pseudoscalars in the final state in addition to the photon, 
hence possibly rendering them accessible to hadronic environments. 

Note that this class of final states also allow
construction of very useful CP-conserving observables, 
in particular, forward-backward asymmetries, 
which are somewhat less clean as therein the SM pollution 
from higher order corrections can be a few percent. 
However, once again, effects of higher order corrections 
cause the asymmetries to be dependent on the specifics
of the hadronic final state and/or the photon energy whereas the SM
pollution from the lowest order $H_{\rm eff}$ is non-vanishing in the SM
due to the finite value of the light quark mass and depends only on that. 
Note also that the LO $H_{\rm eff}$ is generically suited for studying
the effects of NP, since the asymmetries caused by NP effects should
also be independent of the photon energy and/or the details of
the hadronic state accompanying the photon. 
Therefore, these observables
should be useful in probing the existence of NP that gives rise to
current structure different from the left-handed electroweak theory
of the SM~\cite{lrmodels,lrsusy,aps}.



 Let us mention very briefly that direct CP asymmetries
in $PV\gamma$ final states that are driven at the quark level
by $b \to d$ are less clean than the ones compared to $b \to s$.
For the $b \to d$ case, CP-violating triple-correlation
asymmetries as well as CP-conserving
(forward-backward) asymmetries may have a SM ``pollution"
of $O(\delta_{\rm hoc})$, {\it i.e.} a few percent.


\begin{table}[htbp]
\caption{
     SM ``pollution'' expected in various $PV\gamma$ final states
     in $B_u$, $B_d$ and $B_s$ decays. 
     $\delta_{\rm hoc}$ are estimated in~\cite{grinstein1,grinstein2,matsumori,bz}.
   }
\begin{center}
\begin{ruledtabular}
\begin{tabular}{l|l|l}
         decay-type & P-odd, CP-even asymmetry & P-odd, CP-odd asymmetry \\
\hline
         $b \to s$ & $\delta_{\rm hoc} \lsim O(few \%)$ & $\lambda^2 \delta_{\rm hoc} <1
         \%$ \\
\hline
$b \to d$ & $\delta_{\rm hoc} \lsim O(few \%)$ & $\delta_{\rm hoc} \lsim O(few
         \%) $ \\
\hline
\end{tabular}
\end{ruledtabular}
\end{center}
\label{sym}
\end{table}

Table~\ref{sym} gives a compilation of the expected SM
``pollution'' to various category of $PV\gamma$
final states that we have discussed.

Finally, we remark that since parity-odd triple
correlation asymmetries are essential in these
searches, at the first stage a distinction between
$b$ and $\bar b$ decays is not necessary. The latter of course
becomes imperative only after a positive signal is
seen larger than $\delta_{\rm hoc}$, that is, a few percent.

\section*{Acknowledgements}

We thank Dan Pirjol for useful discussions. 
The work of D.~A. and A.~S. are supported in part by US DOE grant Nos.
DE-FG02-94ER40817 (ISU) and DE-AC02-98CH10886 (BNL).

\end{document}